\documentclass[fleqn,usenatbib]{mnras}
\usepackage[T1]{fontenc}
\usepackage{ae,aecompl}
\usepackage[dvipdfmx]{graphicx}	
\usepackage{multirow}
\usepackage{amsmath}
\usepackage{amssymb}
\usepackage{color,ulem}

\newcommand{\Msun}{\ensuremath{\mathrm{M}_\odot}}
\newcommand{\MP}{\ensuremath{m_{\rm{p}}}}
\newcommand{\ME}{\ensuremath{m_{\rm{e}}}}

\newcommand{\dop}{\ensuremath{\delta_{\rm{D}}}}
\newcommand{\nup}{\ensuremath{\nu_{\rm{p}}}}
\newcommand{\Fp}{\ensuremath{F_{\rm{p}}}}
\newcommand{\nua}{\ensuremath{\nu_{\rm{a}}}}
\newcommand{\num}{\ensuremath{\nu_{\rm{m}}}}

\title[Generalized equipartition method]{Generalized equipartition method from an arbitrary viewing angle}

\author[Matsumoto \& Piran]{Tatsuya Matsumoto$^{1}$\thanks{E-mail: tm3238@columbia.edu} and Tsvi Piran$^{2}$\\
$^{1}$Department of Physics and Columbia Astrophysics Laboratory, Columbia University, Pupin Hall, New York, NY 10027, USA\\
$^{2}$Racah Institute of Physics, Hebrew University, Jerusalem, 91904, Israel\\
}

\pubyear{2021}

\begin{document}
\label{firstpage}
\pagerange{\pageref{firstpage}--\pageref{lastpage}}
\maketitle

\begin{abstract}
The equipartition analysis yields estimates of the radius and energy of synchrotron self-absorbed radio sources. Here we generalize this method to relativistic off-axis viewed emitters.
We find that the Lorentz factor $\Gamma$ and the viewing angle $\theta$ cannot be determined independently but become degenerate along a trajectory of minimal energy solutions. The solutions are divided into on-axis and off-axis branches, with the former reproducing the classical analysis. A relativistic source viewed off-axis can be disguised as an apparent Newtonian one. Applying this method to radio observations of several tidal disruption events (TDEs), we find that the radio flare of AT 2018hyz, which was observed a few years after the optical discovery, could have been produced by a relativistic off-axis jet with a kinetic energy of $\sim10^{53}\,\rm erg$ that was launched around the time of discovery.
\end{abstract}


\section{Introduction}
The equipartition method gives estimates of the radius and energy of a radio source showing a synchrotron self-absorbed spectrum (\citealt{Pacholczyk1970,Scott&Readhead1977,Chevalier1998,BarniolDuran+2013}, hereafter BNP13, see also e.g., \citealt{Zdziarski2014b,Petropoulou&Dermer2016} for the case without self-absorption). This method is based on the fact that the total energy of the emitting non-thermal electrons and magnetic field has a very narrow minimum as a function of the outflow's radius. At the minimum, both energies are comparable, and even a tiny deviation from the radius significantly increases the energy. While the original equipartition method applies to only non-relativistic radio sources, BNP13 extended it for relativistic sources moving toward an observer (on-axis emitter). Here we define the ``on-axis'' configuration in which the observer is located within the beaming cone whose half-opening angle is given by $\theta_{\rm b}\simeq1/\Gamma$ of the source. Their extended analysis has been applied to bright radio sources likely powered by a relativistic jet in tidal disruption events (TDEs, \citealt{BarniolDuran&Piran2013,Eftekhari+2018}).

When a radio source is a relativistic and collimated outflow, like a jet, its appearance drastically varies depending on a viewing angle. In particular, the outflow travels in a different direction from an observer's line of sight. Initially, the emission is strongly suppressed to the observer because of the relativistic de-beaming effect. As discussed for gamma-ray bursts (GRBs, \citealt{Granot+2002,Rossi+2002,Totani&Panaitescu2002}), the light curve peaks at a later time when the jet decelerates, and the observer enters its beaming cone (becoming from off to on-axis). This naturally explains the afterglow light curve of GRB 170817A, which we observed at $\simeq0.2\,\rm rad$ from the jet axis \citep[e.g., see ][and references therein]{Margutti&Chornock2021}.

More recently, radio followups for optical TDEs have revealed delayed radio flares \citep{Horesh+2021,Horesh+2021b,Sfaradi+2022,Cendes+2022b,Perlman+2022}. In contrast to typical radio flares following optical TDEs \citep{Alexander+2020}, these delayed events appear a few years after the optical discovery, and some of them are still in a brightening phase \citep{Horesh+2021,Cendes+2022b}. One of the possible origins of the delayed radio flares is an off-axis jet. However, previous analyses were carried out in the assumption of the Newtonian or on-axis case and cannot discuss the possibility of the off-axis jet scenario.

Motivated by the recent detection of delayed radio flares in TDEs, we further extend the equipartition method for an arbitrary viewing angle, i.e., both on and off-axis cases. This paper is organized as follows. In \S \ref{sec:method}, the equipartition method of BNP13 is generalized for arbitrary viewing angle observers. Using the generalized analysis, we develop a new formalism in \S \ref{sec:implication} and apply it to several TDEs whose radio signals are possibly produced by off-axis emission in \S \ref{sec:application}. Finally, we summarize the method and our findings in \S \ref{sec:summary}.

\section{Generalization of equipartition method}\label{sec:method}
\begin{figure}
\begin{center}
\includegraphics[width=85mm, angle=0]{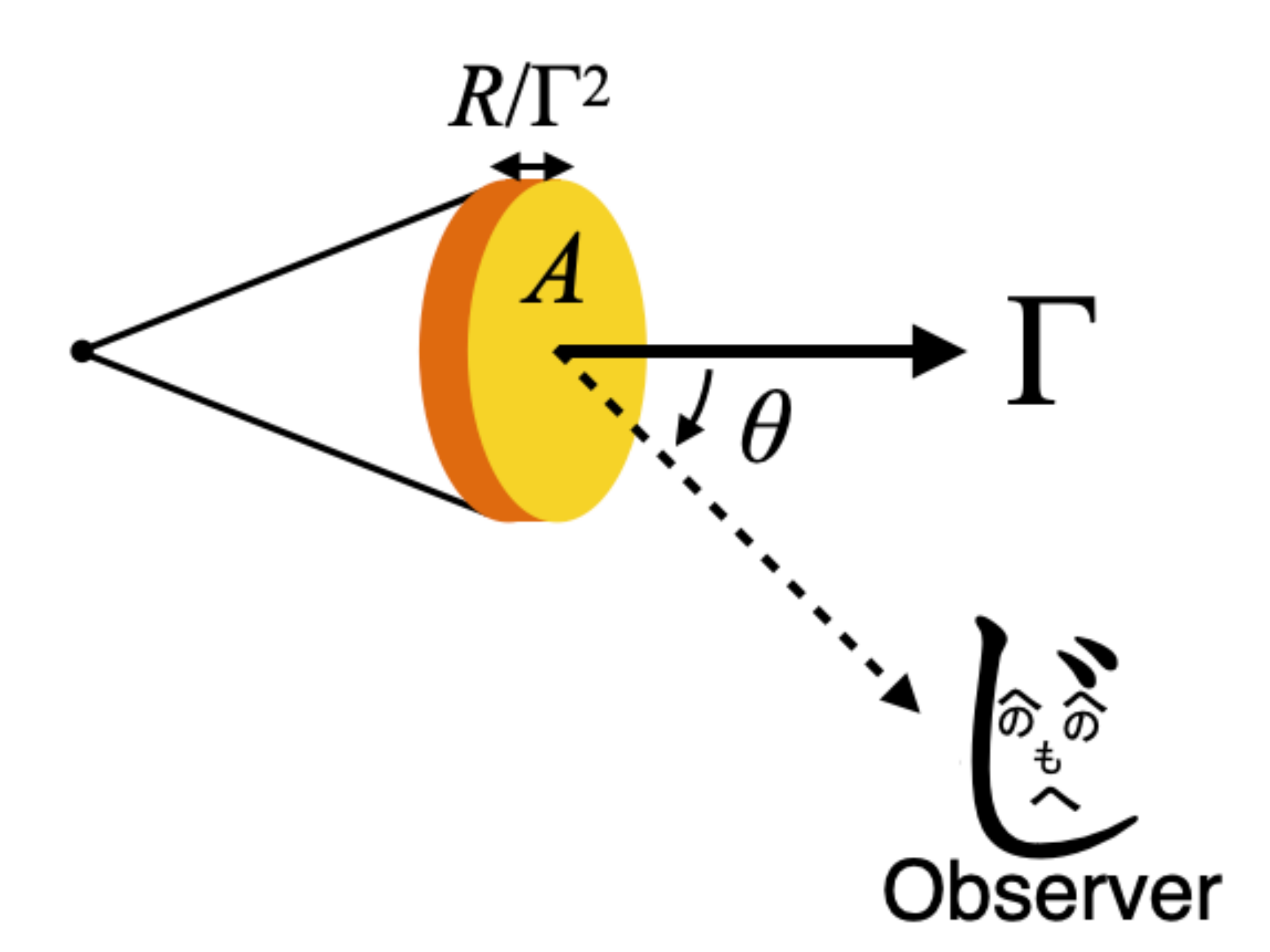}
\caption{A schematic picture. A radio-emitting region is moving at a Lorentz factor $\Gamma$ whose direction of motion is away from the observer's line of sight, $\theta$. The emitting region has an emitting area $A$ and volume of $V$.}
\label{fig:picture}
\end{center}
\end{figure}

Consider a relativistically moving radio source with a Lorentz factor $\Gamma$. The source is at a distance $R$ from the center of an explosion. An observer whose line of sight has an angle of $\theta$ from the source's direction of motion detects its radio-synchrotron emission characterized by a peak frequency $\nup$ and flux density $\Fp$ (see Fig. \ref{fig:picture}).
The relativistic radio source could have an angular structure (i.e., the Lorentz factor and energy vary for different angles). However, due to the strong dependence of the beaming on the Lorentz factor, a small region of order $\pi /\Gamma^2$, dominates the observed emission. Therefore, the radio emission site can be characterized by a single Lorentz factor and regarded as a small patch \citep{Matsumoto+2019,Matsumoto+2019b,Ioka&Nakamura2019}.

The observed quantities are translated from the quantities in the rest frame via the relativistic Doppler factor:
\begin{align}
\dop=\frac{1}{\Gamma\left(1-\beta\cos\theta\right)}\ ,
	\label{eq:dop}
\end{align}
where $\beta\equiv\sqrt{1-1/\Gamma^2}$ is the source velocity normalized by the speed of light $c$.
Note that for a source moving precisely towards the observer ($\theta=0$), the Doppler factor becomes $\dop = 2\Gamma$.
However, BNP13 \citep[following][]{Sari+1998} approximated it as $\dop\simeq\Gamma$ to reflect the fact that the average $\dop$ is lower than $2\Gamma$.\footnote{Averaging the Doppler factor over the beaming cone gives $\left<\dop\right>={\int_0^{1/\Gamma} d\theta\sin\theta\dop}/{(1-\cos\theta)}\simeq(2\ln2)\,\Gamma\simeq1.4\,\Gamma$ for $\Gamma\gg1$ and $\theta\ll1$.} In this paper, we use an exact value of $\dop$ for a given angle to see the off-axis effect.
This treatment leads to some differences in numerical factors between our results at the limit of $\theta=0$ and those of BNP13.\footnote{The exact differences between our equations and those of BNP13 are summarized as follows: Eqs.~\eqref{eq:nup}, \eqref{eq:Fp thin}, \eqref{eq:Fp thick}, and \eqref{eq:B} are twice larger than corresponding Eqs.~(10), (11), (13), and (16) of BNP13 in the limit of $\theta=0$. Eq.~\eqref{eq:gamma_e} is twice smaller than Eq.~(14), Eq.~\eqref{eq:number} is 4 times smaller than Eq.~(15), Eq.~\eqref{eq:E_e} is eight times smaller than Eq.~(17), and Eq.~\eqref{eq:E_B} is 4 times larger than Eq.~(18) of BNP13.}

The observed peak frequency is given by the Doppler-boosted (and redshifted) synchrotron frequency:
\begin{align}
\nu_{\rm p}=\frac{\dop q_{\rm e}B\gamma_{\rm e}^2}{2\pi \ME c (1+z)}\ ,
	\label{eq:nup}
\end{align}
where $q_{\rm e}$ is the elementary charge, $B$ is the magnetic field (at the source rest frame), $\gamma_{\rm e}$ is the Lorentz factor of electrons producing the radio peak, $\ME$ is the electron mass, and $z$ is the redshift to the source.

Two expressions give the peak flux density for optically thin and thick regimes (we describe a more detailed derivation in Appendix \ref{sec:append}).
In the optically thin regime,\footnote{Throughout this paper, we assume that the emission is produced by non-thermal electrons with a power-law energy distribution ($dn/d\gamma\propto\gamma^{-p}$)  in a single zone. Therefore, the spectral index in the optically thin regime should be smaller than $-0.5$ so that the power-law index $p>2$.} the flux density is just given by the flux of a single electron with the Lorentz factor $\gamma_{\rm e}$ multiplied by the number of emitting electrons $N_{\rm e}$:
\begin{align}
F_{\rm p}=\frac{(1+z)\dop^3\sqrt{3}q_{\rm e}^3BN_{\rm e}}{4\pi d_{\rm L}^2\ME c^2}\ ,
	\label{eq:Fp thin}
\end{align}
where $d_{\rm L}$ is the luminosity distance to the source.
We estimate the peak flux by the self-absorbed spectrum in the optically thick regime.
There are potentially two cases depending on the ratio between self-absorption frequency $\nua$ and the characteristic synchrotron frequency $\num$ (corresponding to the emitting electrons with the least energy, see e.g., \citealt{Sari+1998}).
In the case of $\nua>\num$, the flux at $\num$ is suppressed by self-absorption and the radio flux peaks at $\nua$.
The peak flux is given by the Rayleigh-Jeans spectrum:
\begin{align}
F_{\rm p}\simeq\frac{(1+z)^3\dop2\ME \gamma_{\rm e} \nu_{\rm p}^2A}{d_{\rm L}^2}\ ,
	\label{eq:Fp thick1}
\end{align}
where $A$ is the surface area of the emitting region.
In the opposite case of $\num>\nua$, the flux peaks at $\num$ which is obtained by extending the self-absorbed spectrum:
\begin{align}
F_{\rm p}\simeq\frac{(1+z)^3\dop2\ME \gamma_{\rm e}\nu_{\rm a}^2A}{d_{\rm L}^2}\left(\frac{\nu_{\rm p}}{\nua}\right)^{1/3}\ .
\end{align}
Combining the two cases, the peak flux is given by
\begin{align}
F_{\rm p}&=\frac{(1+z)^3\dop2\ME \gamma_{\rm e} \nu_{\rm p}^2A}{3d_{\rm L}^2}\eta^{1/3}\ ,
	\label{eq:Fp thick}\\
\eta&\equiv\begin{cases}
1&;\,\nu_{\rm a}>\nu_{\rm m}\ ,\\
\nu_{\rm m}/\nu_{\rm a}&;\,\nu_{\rm a}<\nu_{\rm m}\ ,
\end{cases}
\end{align}
where following BDP13 we introduced a numerical factor 3 in the denominator of Eq.~\eqref{eq:Fp thick}.

We solve Eqs.~\eqref{eq:nup}, \eqref{eq:Fp thin}, \eqref{eq:Fp thick} to obtain $\gamma_{\rm e}$, $N_{\rm e}$, and $B$:
\begin{align}
\gamma_{\rm e}&=\frac{3F_{\rm p}d_{\rm L}^2\eta^{5/3} \Gamma^2}{2\pi \nu_{\rm p}^2(1+z)^3\ME f_{\rm A}R^2\dop}
	\label{eq:gamma_e}\\
&\simeq5.2\times10^{2}\,\left[\frac{F_{\rm p,mJy}d_{\rm L,28}^2\eta^{5/3}}{\nu_{\rm p,10}(1+z)^3}\right]\frac{\Gamma^2}{f_{\rm A}R_{17}^2\dop} \ ,
	\nonumber\\
N_{\rm e}&=\frac{9cF_{\rm p}^3d_{\rm L}^6\eta^{10/3}\Gamma^4}{2\sqrt{3}\pi^2q_{\rm e}^2\ME^2\nu_{\rm p}^5(1+z)^8f_{\rm A}^2R^4\dop^4}
	\label{eq:number}\\
&\simeq4.1\times10^{54}\,\left[\frac{F_{\rm p,mJy}^3d_{\rm L,28}^6\eta^{10/3}}{\nu_{\rm p,10}^5(1+z)^8}\right]\frac{\Gamma^4}{f_{\rm A}^2R_{17}^4\dop^4} \ ,
	\nonumber\\
B&=\frac{8\pi^3\ME^3c \nu_{\rm p}^5(1+z)^7f_{\rm A}^2R^4\dop}{9q_{\rm e} F_{\rm p}^2d_{\rm L}^4\eta^{10/3}\Gamma^4}
	\label{eq:B}\\
&\simeq1.3\times10^{-2}{\,\rm G\,}\left[\frac{\nu_{\rm p,10}^5(1+z)^7}{F_{\rm p,mJy}^2d_{\rm L,28}^4\eta^{10/3}}\right]\frac{f_{\rm A}^2R_{17}^4\dop}{\Gamma^4} \ ,
	\nonumber
\end{align}
where we use the convention $Q_x = Q/10^x$ (cgs) except for the flux density $F_{\rm p,mJy}=F_{\rm p}/{\rm mJy}$.
The emitting area is measured in units of a surface area of a sphere with a radius $R$, subtending a solid angle of $\pi/\Gamma^2$. We define an area-filling factor following BNP13:
\begin{equation} 
f_{\rm A}\equiv A/\left(\pi R^2/\Gamma^2\right) \ .
    \label{eq:f_A}
\end{equation}
A volume filling factor is also defined by measuring the emitting volume in units of a typical volume of a relativistic shell, i.e., a shell with a radius $R$, width $R/\Gamma^2$, and solid angle of $\pi/\Gamma^2$:
\begin{equation} 
f_{\rm V}=V/\left(\pi R^3/\Gamma^4\right) \ .
    \label{eq:f_V}
\end{equation}
These definitions of filling factors\footnote{While BNP13 originally introduced these geometrical factors for the relativistic case, they turned out to be significant in the Newtonian case as well \citep{Yalinewich+2019b,Matsumoto&Piran2021b}.} are useful and relevant for the emitting source at a radius $R$ from the explosion center moving at a Lorentz factor $\Gamma$. In Appendix \ref{sec:append2}, we discuss the case in which the outflow has a fixed half-opening angle $\theta_{\rm j}$ and hence $f_{\rm A}=f_{\rm V}=(\Gamma\theta_{\rm j})^2$.

Once the number of electrons is obtained (Eq.~\ref{eq:number}), we estimate the number density of the circum-nuclear medium (CNM):
\begin{align}
n_{\rm e}&\simeq\frac{N_{\rm e}}{\frac{\Omega}{3}R^3}=\frac{27cF_{\rm p}^3d_{\rm L}^6\eta^{10/3}\Gamma^4}{2\sqrt{3}\pi^3q_{\rm e}^2\ME^2\nu_{\rm p}^5(1+z)^8f_{\rm A}^2f_{\Omega}R^4\dop^4}
	\label{eq:density}\\
&\simeq3.9\times10^{3}{\,\rm cm^{-3}\,}\left[\frac{F_{\rm p,mJy}^3d_{\rm L,28}^6\eta^{10/3}}{\nu_{\rm p,10}^5(1+z)^8}\right]\frac{\Gamma^6}{f_{\rm A}^2f_{\Omega}R_{17}^7\dop^4} \ ,
	\nonumber
\end{align}
where $\Omega$ is the solid angle subtended by the outflow, and we define the filling factor $f_{\Omega}\equiv\Omega/(\pi/\Gamma^2)$ following Eqs.~\eqref{eq:f_A} and \eqref{eq:f_V}. Note this number density is a lower limit because all electrons are assumed to be accelerated to relativistic energies \cite[e.g., see][]{Matsumoto&Piran2021b}.

The emitting electron's energy and the magnetic field's energy in the lab frame are calculated by
\begin{align}
E_{\rm e}&=\ME c^2 \gamma_{\rm e} \Gamma N_{\rm e}=\frac{27c^3F_{\rm p}^4d_{\rm L}^8\eta^5\Gamma^7}{4\sqrt{3}\pi^3 q_{\rm e}^2\ME^2\nu_{\rm p}^7(1+z)^{11}f_{\rm A}^3R^6\dop^5}
	\label{eq:E_e}\\
&\simeq1.8\times10^{51}{\,\rm erg\,}\left[\frac{F_{\rm p,mJy}^4d_{\rm L,28}^8\eta^{5}}{\nu_{\rm p,10}^{7}(1+z)^{11}}\right]\frac{\Gamma^7}{f_{\rm A}^{3}R_{17}^{6}\dop^5}\ ,
	\nonumber\\
E_{\rm B}&=\frac{B^2}{8\pi}\Gamma^2 V=\frac{8\pi^6\ME^6c^2\nu_{\rm p}^{10}(1+z)^{14}f_{\rm A}^4f_{\rm V}R^{11}\dop^2}{81q_{\rm e}^2F_{\rm p}^4d_{\rm L}^8\eta^{20/3}\Gamma^{10}}
	\label{eq:E_B}\\
&\simeq6.8\times10^{45}{\,\rm erg\,}\left[\frac{\nu_{\rm p,10}^{10}(1+z)^{14}}{F_{\rm p,mJy}^{4}d_{\rm L,28}^{8}\eta^{20/3}}\right]\frac{f_{\rm A}^4f_{\rm V}R_{17}^{11}\dop^2}{\Gamma^{10}} \ .
	\nonumber
\end{align}

The total energy of the emitting electrons and the magnetic field is 
\begin{align}
E_{\rm tot}&=E_{\rm e}+E_{\rm B}=E_{\rm eq}\left[\frac{11}{17}\left(\frac{R}{R_{\rm eq}}\right)^{-6}+\frac{6}{17}\left(\frac{R}{R_{\rm eq}}\right)^{11}\right]\ ,
	\label{eq:E}
\end{align}
where
\begin{align}
R_{\rm eq}&\equiv R_{\rm eq,N}\Gamma\dop^{-7/17}\ ,
    \label{eq:Req}\\
R_{\rm eq,N}&\equiv\biggl(\frac{3^8cF_{\rm p}^8d_{\rm L}^{16}\eta^{35/3}}{2^{4}11\sqrt{3}\pi^9\ME^8\nu_{\rm p}^{17}(1+z)^{25}f_{\rm A}^{7}f_{\rm V}}\biggl)^{1/17}
	\label{eq:ReqN}\\
&\simeq1.9\times10^{17}{\,\rm cm\,}\left[\frac{F_{\rm p,mJy}^{\frac{8}{17}}d_{\rm L,28}^{\frac{16}{17}}\eta^{\frac{35}{51}}}{\nu_{\rm p,10}(1+z)^{\frac{25}{17}}}\right]f_{\rm A}^{-\frac{7}{17}}f_{\rm V}^{-\frac{1}{17}} \ , 
	\nonumber
\end{align}
and 
\begin{align}
E_{\rm eq}&=E_{\rm eq,N}\Gamma\dop^{-43/17}\ ,
    \label{eq:Eeq}\\
E_{\rm eq,N}&\equiv\left(\frac{17^{17}\pi^3c^{45}\ME^{14}F_{\rm p}^{20}d_{\rm L}^{40}\eta^{15}f_{\rm V}^6}{2^{10}3^211^{11}\sqrt{3}q_{\rm e}^{34}\nu_{\rm p}^{17}(1+z)^{37}f_{\rm A}^{9}}\right)^{1/17}
	\label{eq:EeqN}\\
&\simeq6.2\times10^{49}{\,\rm erg\,}\left[\frac{F_{\rm p,mJy}^{\frac{20}{17}}d_{\rm L,28}^{\frac{40}{17}}\eta^{\frac{15}{17}}}{\nu_{\rm p,10}(1+z)^{\frac{37}{17}}}\right]f_{\rm A}^{-\frac{9}{17}}f_{\rm V}^{\frac{6}{17}}\ .
	\nonumber
\end{align}
Here we factored out the relativistic corrections in Eqs.~\eqref{eq:Req} and \eqref{eq:Eeq} and define the Newtonian equipartition radius and energy by Eqs.~\eqref{eq:ReqN} and \eqref{eq:EeqN}, which are determined by the observables and the geometrical parameters $f_{\rm A}$ and $f_{\rm V}$.

We emphasize that the total energy in Eq.~\eqref{eq:E} contains only emitting electrons and magnetic field energies. The contribution of non-emitting electrons and baryons increases the energy. We refer the readers to BNP13 for such  extensions. However, these corrections do not change the qualitative nature of the solution.

Normalizing the variables to the Newtonian quantities, we define 
\begin{align} 
r\equiv R/R_{\rm eq,N}\,\,\,{\rm and}\,\,\,e\equiv E/E_{\rm eq,N}\ ,
\end{align}
and now rewrite Eq.~\eqref{eq:E} as: 
\begin{align}
e(r,\Gamma,\theta)=\Gamma\dop^{-43/17}\left[\frac{11}{17}\left(\frac{r}{\Gamma\dop^{-7/17}}\right)^{-6}+\frac{6}{17}\left(\frac{r}{\Gamma\dop^{-7/17}}\right)^{11}\right]\ .
	\label{eq:E2}
\end{align}
The relativistic effects are now seen clearly in the appearance of $\Gamma$ and $\dop$. Note, however, that they also appear indirectly in the definitions of geometrical factors $f_{\rm A}$ and $f_{\rm V}$.

\section{Minimal energy}\label{sec:implication}
In the Newtonian equipartition analysis ($\Gamma=1$ and $\dop=1$), the total energy depends only on the radius, and it is minimized at $R_{\rm eq,N}$. As is well known, a slight deviation from $R_{\rm eq,N}$ increases the total energy by orders of magnitude, and hence the actual radius is     $\approx R\simeq R_{\rm eq,N}$ and the energy is not much larger than $E\simeq E_{\rm eq,N}$.
In contrast, for the relativistic regime, with the additional two variables $\Gamma$ and $\theta$, the energy no longer has a global minimum. An additional consideration is required to determine the radius and energy.

\begin{figure}
\begin{center}
\includegraphics[width=85mm, angle=0]{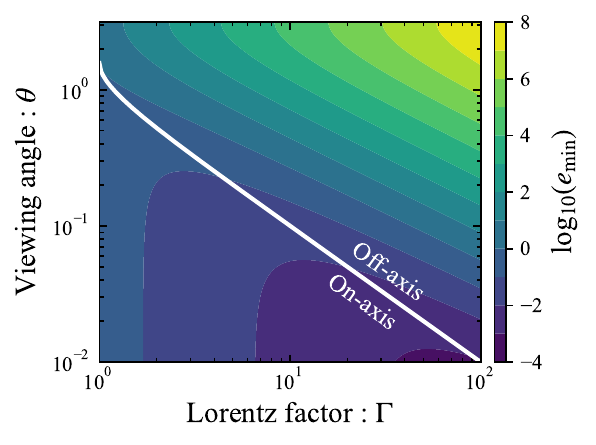}
\caption{The distribution of the minimal normalized energy (Eq.~\ref{eq:e_min}) for different Lorentz factors and viewing angles. The on and off-axis regions are divided by the white line $\theta \simeq 1/\Gamma$ (see Eq.~\ref{eq:beaming cone}). For given observations, the energy is smaller for on-axis observers than off-axis ones, and its value decreases for larger Lorentz factors.}
\label{fig:e_min}
\end{center}
\end{figure}

\begin{figure*}
\begin{center}
\includegraphics[width=185mm, angle=0]{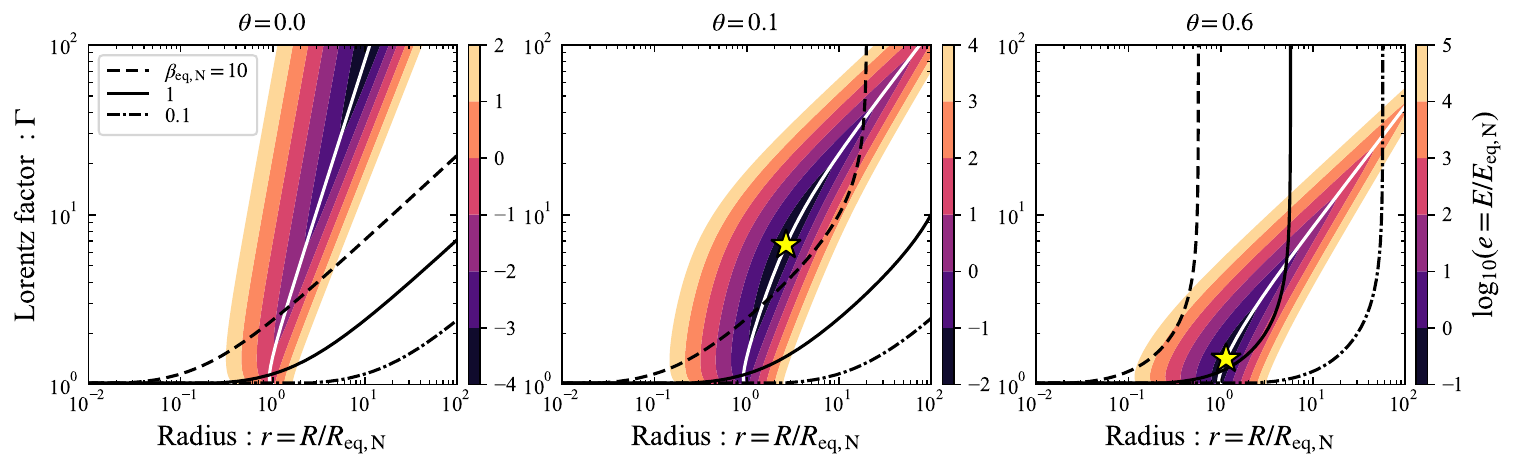}
\caption{Distributions of the total energy (Eq.~\ref{eq:E2}) as a function of radius and Lorentz factor for different viewing angles of $\theta=0$, 0.1, and 0.6 (left to right). White curves represent the radius minimizing the energy for each Lorentz factor and angle (Eq.~\ref{eq:r}). Black curves show the contours for fixed $\beta_{\rm eq,N}=R_{\rm eq,N}(1+z)/(ct)=10$ (dashed), 1 (solid), and 0.1 (dash-dotted). For the middle and right panels, star points denote the location where the energy is minimized for the given angle.}
\label{fig:dist}
\end{center}
\end{figure*}

\begin{figure}
\begin{center}
\includegraphics[width=85mm, angle=0]{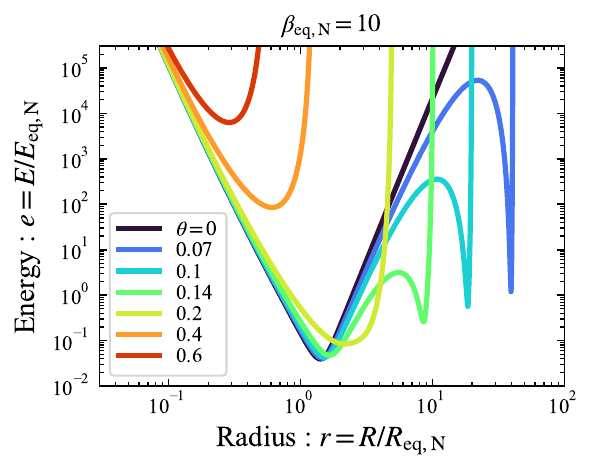}
\caption{Total energy as a function of radius. This is the variation of energy along black dashed curves (Eq.~\ref{eq:r}) in Fig.~\ref{fig:dist} for different $\theta$. We fix the ratio $\beta_{\rm eq,N}=10$ while different values of $\beta_{\rm eq,N}$ give a similar behavior qualitatively. For $\theta<\theta_{\rm c}\simeq0.2$ (Eq.~\ref{eq:theta_c}), the energy has two minima corresponding to on (smaller radius) and off-axis (larger radius) solutions.}
\label{fig:re}
\end{center}
\end{figure}

Still, even in the relativistic case, the energy has a very sharp minimum as a function of $r$,
\begin{equation}
e_{\rm min}(\Gamma,\theta) =\Gamma\dop^{-43/17}\ ,
    \label{eq:e_min}
\end{equation} 
at
\begin{align}
r=\Gamma\dop^{-7/17}\ .
    \label{eq:req}
\end{align}
Fig.~\ref{fig:e_min} depicts the minimal energy $e_{\rm min}$ as a function of $\Gamma$ and $\theta$.
For a fixed Lorentz factor, the energy is always minimized for an exactly on-axis configuration ($\theta=0$), and it does not vary so much as long as the emission is viewed from on-axis $\theta < 1/\Gamma$. In other words, off-axis sources require more energy than on-axis ones, and the energy increases by many orders of magnitude as $\theta$ increases. For an on-axis observer, the minimal energy decreases quickly, $e_{\rm min}\propto \Gamma^{-26/17}$, as the source becomes more relativistic (BNP13, and see the left panel of Fig.~\ref{fig:dist}).
For a fixed viewing angle, the energy has a minimum around the boundary between the on and off-axis.

The behavior of $e_{\rm min}$ is explained by the fact that the emission is amplified by the substantial Doppler boost, and less energy is needed to produce a given observed flux. Note that the geometrical factors $f_{\rm A}$ and $f_{\rm V}$ are assumed to be constant here. Namely, the actual physical size of the emitting region decreases for larger $\Gamma$, which also helps to reduce the minimal energy.\footnote{This does not necessarily mean that the whole energy of the explosion is small. Most of the emission from other regions is beamed away and hidden from the observer at this stage.} 
Regardless of this unphysical assumption of a highly narrow emitter moving at an extremely large Lorentz factor, this result implies that unlike the Newtonian case, there is no global minimum for the energy of the source, and other considerations have to be added to determine the conditions at the source entirely.

The definition of on (off) axis, $\theta<(>)1/\Gamma$ becomes irrelevant for Newtonian sources because $\Gamma\to1$ results in $\theta=1$ as the boundary between the on and off-axis. This is not true because an emission from the source can be seen from any angle. Here, for completeness, we define the size of the beaming cone by an angle within which half of the photons are isotropically emitted in the source rest frame is confined in the lab frame. With this definition, the beaming cone is given by 
\begin{align}
\theta_{\rm b}=\cos^{-1}\beta\ ,
    \label{eq:beaming cone}
\end{align}
which asymptotes to $\theta_{\rm b}\to1/\Gamma$ for $\Gamma\gg1$ restoring the conventional definition of the beaming angle. Hereafter we use this definition to depict the boundary between on and off-axes in figures.  

Fig.~\ref{fig:dist} depicts the distribution of the normalized energy, $e(r,\Gamma,\theta)$ (Eq.~\ref{eq:E2}) in the $(r,\Gamma)$ plane for different viewing angles.
As a function of $r$, the energy takes a minimal value for $r=\Gamma\dop^{-7/17}$ (white curve on each panel).
As we have already seen that in the relativistic case, due to the additional dependence on $\Gamma$ and $\theta$, the energy (Eq.~\ref{eq:E2}) no longer has a global minimum. Indeed for $\theta=0$ (the left panel in Fig.~\ref{fig:dist}), the energy $e$ is arbitrarily decreased by increasing $\Gamma$.
For finite angles, $\theta\neq 0$, the energy has a minimal value at a point on $r=\Gamma\dop^{-7/17}$ (stars in the middle and right panels in Fig.~\ref{fig:dist}).
However, the variation of the total energy along the trajectory of $r=\Gamma\dop^{-7/17}$ is much milder than the case in which $r$ deviates from $\Gamma\dop^{-7/17}$. Therefore, even for a finite angle case, requiring minimal energy, as in the Newtonian case, is insufficient in the relativistic case.

Following BNP13 we introduce an additional condition on the three variables.
When the moment of the explosion (or equivalently, the launch of the outflow) is observationally identified, the radius and the observation time $t$ (measured since the explosion in the observer frame) are related by 
\begin{align}
t=\frac{(1+z)R}{c\beta} (1-\beta \cos \theta) \ .
    \label{eq:t}
\end{align}
This gives a second relation between the three variables: 
\begin{align}
r&=\left(\frac{\beta}{\beta_{\rm eq,N}}\right) \Gamma\dop\ ,
	\label{eq:r}\\
\beta_{\rm eq,N}&\equiv\frac{(1+z)R_{\rm eq,N}}{ct}
    \label{eq:beta}\\
&\simeq0.73\,\left[\frac{F_{\rm p,mJy}^{\frac{8}{17}}d_{\rm L,28}^{\frac{16}{17}}\eta^{\frac{35}{51}}}{\nu_{\rm p,10}(1+z)^{\frac{8}{17}}}\left(\frac{t}{100\,\rm day}\right)^{-1}\right]f_{\rm A}^{-{7}/{17}}f_{\rm V}^{-{1}/{17}} \ .
	\nonumber
\end{align}
The parameter $\beta_{\rm eq,N}$ describes an apparent velocity of the emitting source. Notably, $\beta_{\rm eq,N}<1$ suggests that the source is Newtonian. However, we will show this is not necessarily the case.

Black lines in Fig.~\ref{fig:dist} show contours of Eq.~\eqref{eq:r} for $\beta_{\rm eq,N}=10$, 1, and 0.1.
The radius and the Lorentz factor are restricted along a curve for a given $\beta_{\rm eq,N}$ by an observation (with fixed geometrical parameters $f_{\rm A}$ and $f_{\rm V}$).
For $\theta=0$, this curve always intersects the curve of $r=\Gamma\dop^{-7/17}$ at a single point, which gives a unique estimate of the radius and corresponding Lorentz factor (BNP13).
In contrast, these curves may intersect twice or never for  $\theta \ne 0$. This is because Eq.~\eqref{eq:r} asymptotes to $r\simeq2/(\beta_{\rm eq,N}\theta^2)$ for $\Gamma\gtrsim1/\theta$ (the observed time caps the outflow's radial distance).
When there are two intersections, there are two corresponding minimal energies. When there is no intersection, the energy takes a minimum at a smaller radius than $r=\Gamma\dop^{-7/17}$, and its value is typically much larger than that obtained on this line.
Fig.~\ref{fig:re} shows the behavior of the energy (Eq.~\ref{eq:E2}) along the radius given by Eq.~\eqref{eq:r} for different viewing angles. For smaller (but non-zero) viewing angles, the energy has two minimal values, as shown in Fig.~\ref{fig:dist}.

\begin{figure*}
\begin{center}
\includegraphics[width=185mm, angle=0]{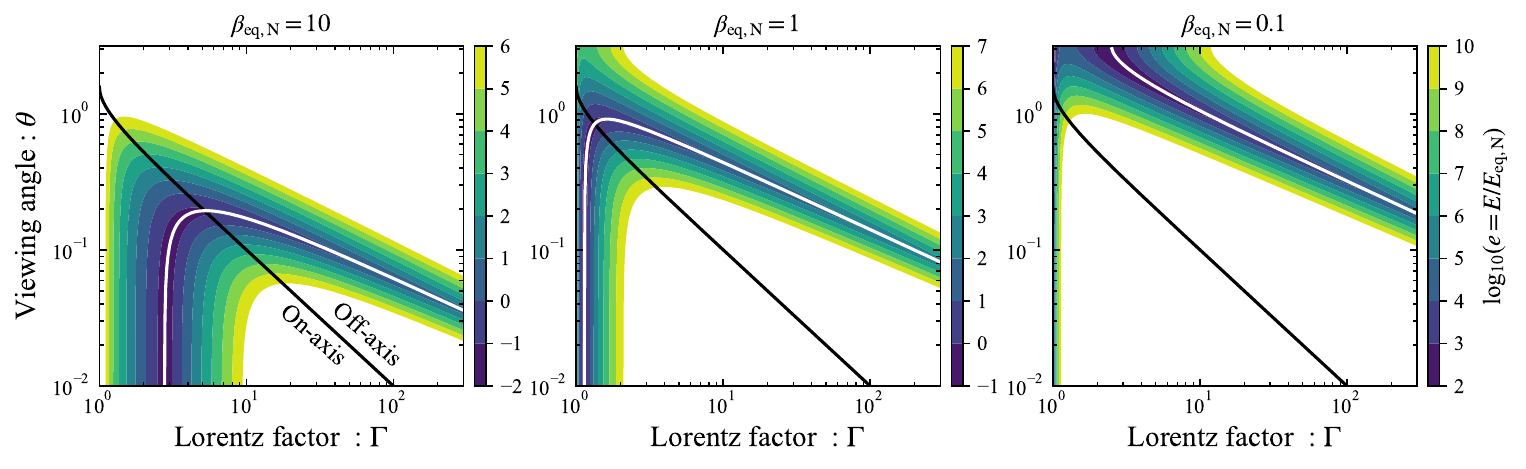}
\caption{Distributions of the total energy $e$ (Eq.~\ref{eq:E2}) with the condition of Eq.~\eqref{eq:r}, as a function of Lorentz factor and viewing angle for different $\beta_{\rm eq,N}=10$, 1, and 0.1 (left to right). The on and off-axis regions are divided by the black line $\theta\simeq1/\Gamma$. White curves give a sequence of minimal energy (Eq.~\ref{eq:dop_result}) and possible parameter sets of a radio-emitting source with a given set of observables.}
\label{fig:dist2}
\end{center}
\end{figure*}

For given $\theta$ and $\beta_{\rm eq,N}$, we derive the minimal energies and corresponding Lorentz factor and radii, which give the intersections of the white and black curves in Fig.~\ref{fig:dist}. Combining Eqs.~\eqref{eq:req} and \eqref{eq:r}, we find the Doppler factor is constrained
\begin{align}
\dop=\left(\frac{\beta}{\beta_{\rm eq,N}}\right)^{-17/24}\ .
	\label{eq:dop_result}
\end{align}
Given a viewing angle, we solve Eq.~\eqref{eq:dop_result} for the Lorentz factor and calculate the corresponding minimal energy and radius:
\begin{align}
e_{\rm min}&=\left(\frac{\beta}{\beta_{\rm eq,N}}\right)^{43/24}\Gamma\ ,
    \label{eq:e_result}\\
r&=\left(\frac{\beta}{\beta_{\rm eq,N}}\right)^{7/24}\Gamma\ .
    \label{eq:r_result}
\end{align}

In the limit of $\Gamma\gg1$ and $\theta\ll1$, we can obtain more explicit results instead  of Eqs.~\eqref{eq:dop_result}-\eqref{eq:r_result}. In this limit, the Doppler factor is approximated by
\begin{align}
\dop\simeq\frac{2\Gamma}{1+(\Gamma\theta)^2}\ ,
    \label{eq:dop2}
\end{align}
and we can rewrite Eq.~\eqref{eq:dop_result} as
\begin{align}
(\Gamma\theta)^2-2\beta_{\rm eq,N}^{-17/24}\Gamma+1\simeq0\ .
	\label{eq:theta}
\end{align}
For $\theta=0$, the Lorentz factor and corresponding minimal energy and radius are given by
\begin{align}
\Gamma_{\rm on}&\simeq\frac{\beta_{\rm eq,N}^{17/24}}{2}\ ,  
    \label{eq:gam_on}\\
e_{\rm min,on}&\simeq\frac{1}{2\beta_{\rm eq,N}^{13/12}}\ ,   
    \label{eq:e_min_on}\\
r_{\rm on}&\simeq\frac{\beta_{\rm eq,N}^{5/12}}{2}\ .
    \label{eq:r_on}
\end{align}
These equations are basically the same as those given by BNP13.
Here we implicitly assume $\beta_{\rm eq,N}\gtrsim1$ so that we have $\Gamma_{\rm on}\gtrsim1$. In the opposite case of $\beta_{\rm eq,N}<1$, we have the same results as the Newtonian equipartition method: $\Gamma_{\rm on}\simeq e_{\rm min,on}\simeq r_{\rm on}\simeq1$.

Eq.~\eqref{eq:theta} has no solution when the viewing angle is larger than a critical value,
\begin{align}
\theta_{\rm c}=\beta_{\rm eq,N}^{-17/24}\ .
	\label{eq:theta_c}
\end{align} 
For $0< \theta < \theta_{\rm c}$
there are two solutions:
\begin{align}
\Gamma\simeq\frac{1\pm\sqrt{1-\beta_{\rm eq,N}^{17/12}\theta^2}}{\beta_{\rm eq,N}^{17/24}\theta^2}\ .
\end{align}
The negative sign corresponds to the on-axis solution.  It converges to Eq.~\eqref{eq:gam_on} in the $\theta\to 0$ limit.
The positive sign corresponds to the off-axis configuration.\footnote{For this solution,  $\Gamma\theta>1$ under the condition of Eq.~\eqref{eq:theta_c}.} Therefore, this branch is  simply described by taking the limit $\Gamma\theta\gg1$ in Eq.~\eqref{eq:theta}: 
\begin{align}
\Gamma_{\rm off}&\simeq\frac{2}{\beta_{\rm eq,N}^{17/24}\theta^2}\ ,
    \label{eq:gam_off}\\
e_{\rm min,off}&\simeq\frac{2}{\beta_{\rm eq,N}^{5/2}\theta^2}\ ,
    \label{eq:e_min_off}\\
r_{\rm off}&\simeq\frac{2}{\beta_{\rm eq,N}\theta^2}
    \label{eq:r_off}\ .
\end{align}
The ratios of the Lorentz factors, energies, and radii of on-axis to off-axis solutions are given by:
\begin{align}
\frac{\Gamma_{\rm on}}{\Gamma_{\rm off}}=\frac{e_{\rm min,on}}{e_{\rm min,off}}=\frac{r_{\rm on}}{r_{\rm off}}=\frac{\beta_{\rm eq,N}^{17/12}\theta^2}{4}=\left(\frac{\theta}{2\theta_{\rm c}}\right)^2<1\ .
    \label{eq:on/off}
\end{align} 
where $\theta$ here is for the off-axis solution.
The off-axis branch always has a larger minimal energy and a larger radius than the on-axis ones.

The above equations nicely explain the behavior of the minimal energies shown in Fig.~\ref{fig:re}.
For  $\theta<\theta_{\rm c}\simeq0.2\,(\beta_{\rm eq,N}/10)^{-17/24}$ (Eq.~\ref{eq:theta_c}), the energy has two minima corresponding to the on and off-axis solutions.
The on-axis solution has a lower energy at a smaller radius than the off-axis one. The minimal energy of the on-axis solution is only weakly dependent on the viewing angle. 
For the off-axis solution, the radius is larger, and the emitter is accordingly more relativistic than the on-axis one.
The minimum around this radius is extremely narrow, and a slight deviation from the minimal radius drastically increases the energy.
For  $\theta>\theta_{\rm c}$, the energy has a single minimal value at a radius smaller than that for $\theta<\theta_{\rm c}$. This is because Eqs.~\eqref{eq:req} and \eqref{eq:r} do not hold at the same time. This minimal value of the energy increases rapidly with $\theta$, and typically it is much larger than $e_{\rm min}$ (or $E_{\rm eq}$).

Fig.~\ref{fig:dist2} depicts the distribution of the normalized energy, $e(r,\Gamma,\theta)$ (Eq.~\ref{eq:E2}) in the $(\Gamma,\theta)$ plane for different $\beta_{\rm eq,N}$ under the condition of Eq.~\eqref{eq:r}.
The locus of the minimal energy is given by Eq.~\eqref{eq:dop_result} or approximately described by solving Eq.~\eqref{eq:theta} for $\theta$:
\begin{align}
\theta\simeq\left(\frac{2}{\beta_{\rm eq,N}^{17/24}\Gamma}-\frac{1}{\Gamma^2}\right)^{1/2}\ .
	\label{eq:theta2}
\end{align}
As we have discussed, this locus is divided into on and off-axis branches.  The Lorentz factor of the off-axis branch is always larger than that of the on-axis one (see Eq.~\ref{eq:on/off}). 
Interestingly, a source interpreted as a Newtonian emitter with $\beta_{\rm eq,N}\lesssim1$ can be a relativistic one viewed off-axis (see also the right panel of Fig.~\ref{fig:dist2}). While more energy is required for the relativistic off-axis configuration, there are situations where this is not a problem, and the off-axis solution is the right one.

\begin{figure}
\begin{center}
\includegraphics[width=85mm, angle=0]{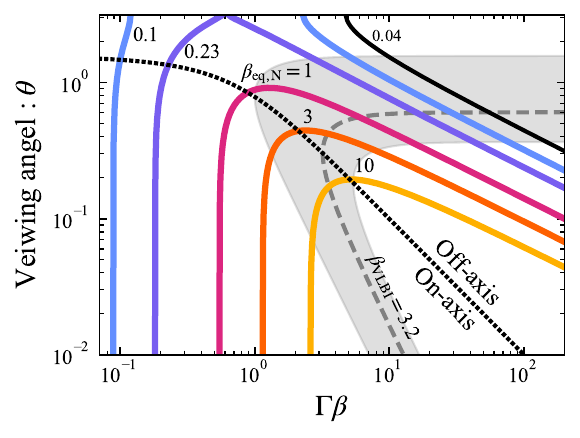}
\caption{Minimal energy trajectories for different values of $\beta_{\rm eq,N}$ in the $(\Gamma \beta, \theta)$ plane.
For $\beta_{\rm eq,N}< 0.23$, the trajectory has discrete Newtonian (on-axis) and relativistic (off-axis) branches. The gray dashed curve and the colored stripe around it denote the contour corresponding to an apparent superluminal velocity $\beta_{\rm VLBI}=3.2\pm 2.2$. The intersection of a trajectory with the stripe describes a unique solution. The values of $\beta_{\rm VLBI}$ and the black solid curve with $\beta_{\rm eq,N}=0.04$ correspond to the observations of AT 2019dsg.}
\label{fig:veq}
\end{center}
\end{figure}

Fig.~\ref{fig:veq} shows a sequence of minimal energy trajectories for different values of $\beta_{\rm eq,N}$ in the ($\Gamma\beta,\theta$) plane. For a given observation with $\beta_{\rm eq,N}$, the Lorentz factor and viewing angle are not determined independently, but they can vary along this trajectory. As expected, for smaller $\beta_{\rm eq,N}$ values, the on-axis four-velocity approaches the apparent velocity $(\Gamma\beta)\to\beta_{\rm eq,N}$. For small values of  $\beta_{\rm eq,N}\lesssim0.23$, the minimal energy trajectory disappears for $\Gamma\beta\sim 1$, and the trajectory is separated into disconnected  Newtonian (on-axis)  and relativistic (off-axis) branches. This may be understood by noting the velocity parameter is related to the radio luminosity, $F_{\rm p} d_{\rm L}^2  \propto \beta_{\rm eq,N}^{17/8}$, and hence a smaller $\beta_{\rm eq,N}$ corresponds to a dim source. However, given the strong sensitivity of radio luminosity on the velocity \citep[e.g.,][]{Nakar&Piran2002,Bruni+2021} if $\Gamma \beta \simeq 1$ the source will be too bright and inconsistent with the observed one.  A large $\theta$ leads to a small Lorentz boost that quenches the observed signal. However, such a solution is strongly off-axis and requires a very large $\Gamma\beta$.

For a single epoch observation that determines  $\beta_{\rm eq,N}$, the Lorentz factor and the viewing angle cannot be determined uniquely as there is a degeneracy along the minimal energy trajectory. However, we can break this degeneracy by adding another observational input. Promising information is an apparent velocity obtained by a VLBI observation. The displacement of the emitting region on the sky plane gives an apparent speed:
\begin{align}
\beta_{\rm VLBI}=\frac{\beta\sin\theta}{(1-\beta\cos\theta)(1+z)}\ .
    \label{eq:beta_vlbi}
\end{align}
In Fig.~\ref{fig:veq}, we show such a trajectory for $\beta_{\rm VLBI}=3.2$ (motivated by the observation of a TDE, see \S\ref{sec:at2019dsg}). It intersects with the minimal energy trajectory, and hence a VLBI observation breaks the degeneracy between $\Gamma$ and $\theta$.

Since the equipartition method gives both the radius and density (Eq.~\ref{eq:density}), it can be used to infer the density profile of galactic nuclear regions \citep[e.g.,][]{BarniolDuran&Piran2013,Zauderer+2013,Alexander+2016,Krolik+2016}. For off-axis observers, the outflow radius increases, and the density profile differs from the on-axis one. By Eqs.~\eqref{eq:density}, \eqref{eq:req}, and \eqref{eq:dop_result}, we find the density at the minimizing radius depends on the parameters as $n_{\rm e}\propto r^{-1}(\beta/\beta_{\rm eq,N})^{13/12}$. Noting that the velocity becomes $\beta\to\beta_{\rm eq,N}$ for an on-axis solution with $\beta_{\rm eq,N}<1$, or $\beta\to1$ otherwise, we obtain the ratio of the densities for the off and on-axis solutions:
\begin{align}
\frac{n_{\rm off}}{n_{\rm on}}\simeq{\rm max}\left[1,\beta_{\rm eq,N}^{-13/12}\right]\left(\frac{r_{\rm off}}{r_{\rm on}}\right)^{-1}\ .
    \label{eq:density_on/off}
\end{align}

\section{Application to observed objects}\label{sec:application}
If the time of the explosion is identified, each observation provides us with the velocity parameter $\beta_{\rm eq,N}$ (Eq.~\ref{eq:beta}) at each epoch. As the Lorentz factor and the viewing angle are degenerate along the minimal energy trajectory given by Eq.~\eqref{eq:dop_result} (see also Fig.~\ref{fig:veq}) we can consider different physical scenarios for the radio source.

\begin{figure}
\begin{center}
\includegraphics[width=85mm, angle=0]{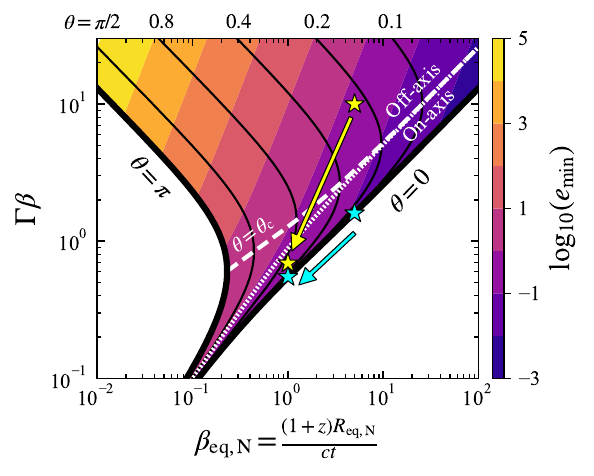}
\caption{The possible range of four-velocity $\Gamma\beta$ given by the minimal energy trajectory for each $\beta_{\rm eq,N}$. The corresponding viewing angle and minimal energy are shown with black and colored contours. White dashed and dotted curves denote the contours of the maximal viewing angle ($\theta_{\rm c}$) and the boundary of on and off-axis configurations ($\theta=\cos^{-1}\beta$). The cyan stars represent the evolution of a decelerating on-axis emitter (the top panel of Fig.~\ref{fig:picture2}). The yellow ones show the case of an emitting region evolving from off to on-axis configurations (the bottom panel of Fig.~\ref{fig:picture2}).}
\label{fig:veq2}
\end{center}
\end{figure}

Fig.~\ref{fig:veq2} depicts the possible range of $\Gamma\beta$ for each value of $\beta_{\rm eq,N}$. For a given $\beta_{\rm eq,N}$, the four-velocity takes the minimal value at $\theta=0$. For larger four-velocities, the viewing angle increases up to the critical angle $\theta_{\rm c}$, which typically coincides with the boundary between the on and off-axis branches for $\beta_{\rm eq,N}\gtrsim1$, and then it decreases along the off-axis branch. When $\beta_{\rm eq,N}$ is smaller than a critical value $\beta_{\rm eq,N}\lesssim0.23$, the possible region of $\Gamma\beta$ is separated into a relativistic and Newtonian regions. In the Newtonian region, the solution is naturally on-axis, and the velocity converges to $\Gamma\beta\to\beta_{\rm eq,N}$.

\begin{figure}
\begin{center}
\includegraphics[width=85mm, angle=0]{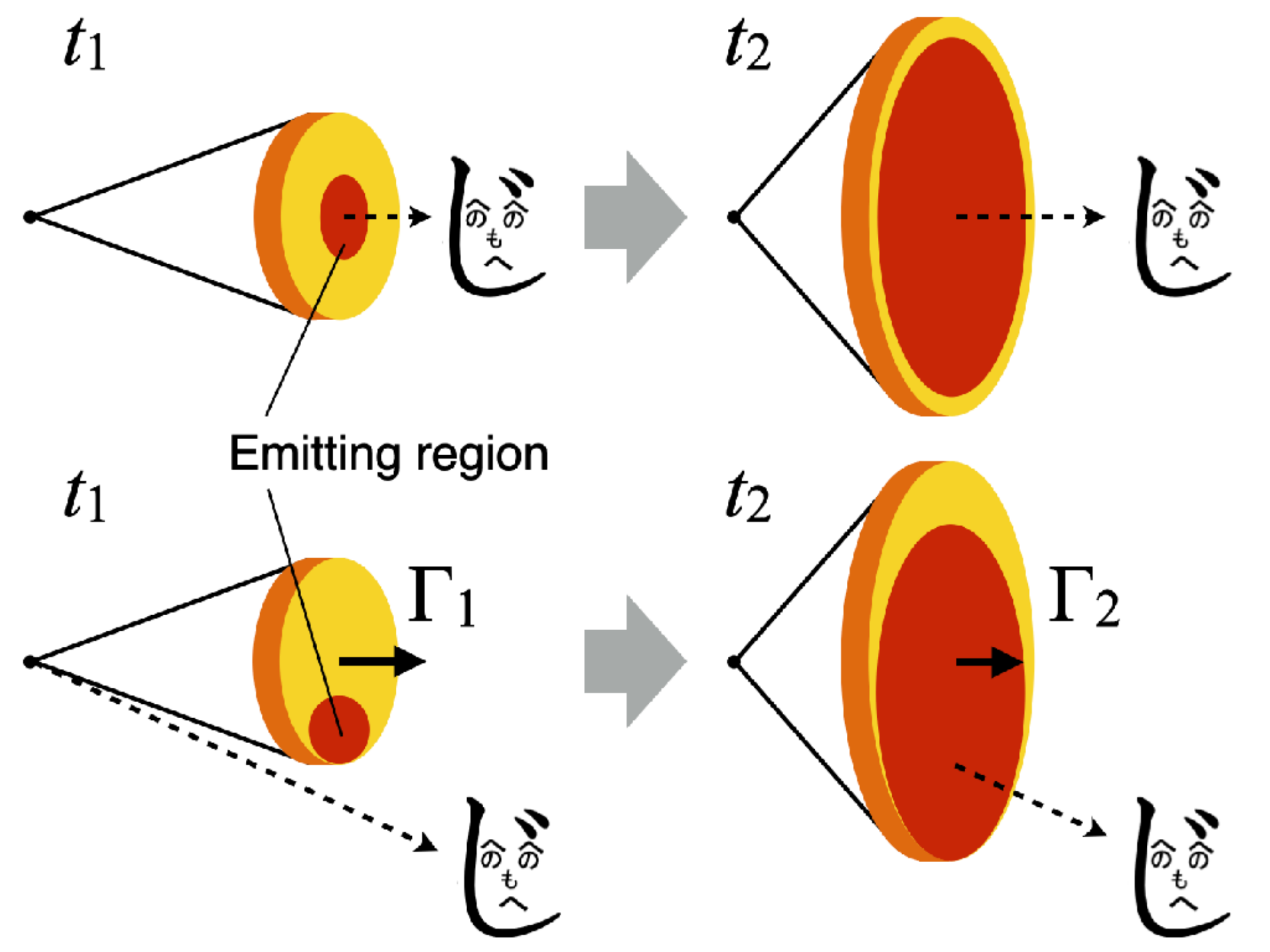}
\caption{A schematic picture of two temporal evolutions of the radio-emitting region corresponding to the cyan and yellow stars in Fig.~\ref{fig:veq2}. ({\bf Top}) The radio emitter is viewed on-axis from epochs $t_1$ to $t_2$ as the emitting region decelerates from $\Gamma_1$ to $\Gamma_2$. The observer's line of sight coincides with the emitter's direction of motion.({\bf Bottom}) The emitter evolves from off to on-axis configurations. The observer is out of the beaming cone at $t_1$ but enters the cone as the radio source decelerates. 
}
\label{fig:picture2}
\end{center}
\end{figure}

We consider observations at two epochs  $t_1$ and $t_2$ measured from the time of the explosion ($t_1<t_2$). As an example, let us consider the case in which the apparent velocity parameter $\beta_{\rm eq,N}$ decreases with time ($\beta_{\rm eq,N}(t_1)>\beta_{\rm eq,N}(t_2)$). Two possible evolution of the radio source are  shown in Fig.~\ref{fig:veq2}. One is an on-axis viewed emitter (cyan stars along the contour of $\theta=0$). The emitting region is moving along the observer's line of sight (as assumed in BNP13). The other is an emitting region evolving from off-axis to on-axis  (yellow stars). This is the case when a relativistic jet is launched in a different direction from an observer's line of sight. Initially, only a small fraction of the jet closer to the line of sight (or ``jet's wing'' if it has an angular structure) dominates the emission. Then, as the jet decelerates, the observer can see the whole jet in an on-axis configuration. In this case, the equipartition energy, which reflects just the energy of the observed region, increases with time as the jet slows down and a larger fraction of the jet comes into view and contributes to the emission. The corresponding schematic pictures for both situations are shown in the top (on-axis) and the bottom (off to on-axis) panels of Fig.~\ref{fig:picture2}.

We turn now to apply our generalized equipartition method to observed objects. We focus on radio emissions from TDEs. A fraction of TDEs is accompanied by a radio flare with a synchrotron self-absorbed spectrum, which is analyzed using the equipartition method \citep[see][for a review]{Alexander+2020}. In the following, we analyze, as an example, two TDEs.

\subsection{AT 2019dsg}\label{sec:at2019dsg}
First, we demonstrate applying the generalized equipartition method to a well-observed event AT 2019dsg \citep{Lee+2020,Cannizzaro+2021,Cendes+2021b,Stein+2021,Mohan+2022}. It has been suggested that AT 2019dsg was associated with a high-energy neutrino \citep{Stein+2021}. 

Previous on-axis equipartition analyses found that the observations are consistent with a Newtonian outflow with $\beta\sim0.1$ launched ten days before the optical discovery  \citep{Cendes+2021b,Matsumoto+2022}. The energy implied was modest of order $\simeq 10^{47}\,\rm erg$. In contrast to these results, some theoretical models for the neutrino emission require a relativistic and more energetic outflow \citep[e.g.,][]{Winter&Lunardini2021}, which can be tested by our analysis. 

\begin{figure}
\begin{center}
\includegraphics[width=85mm, angle=0]{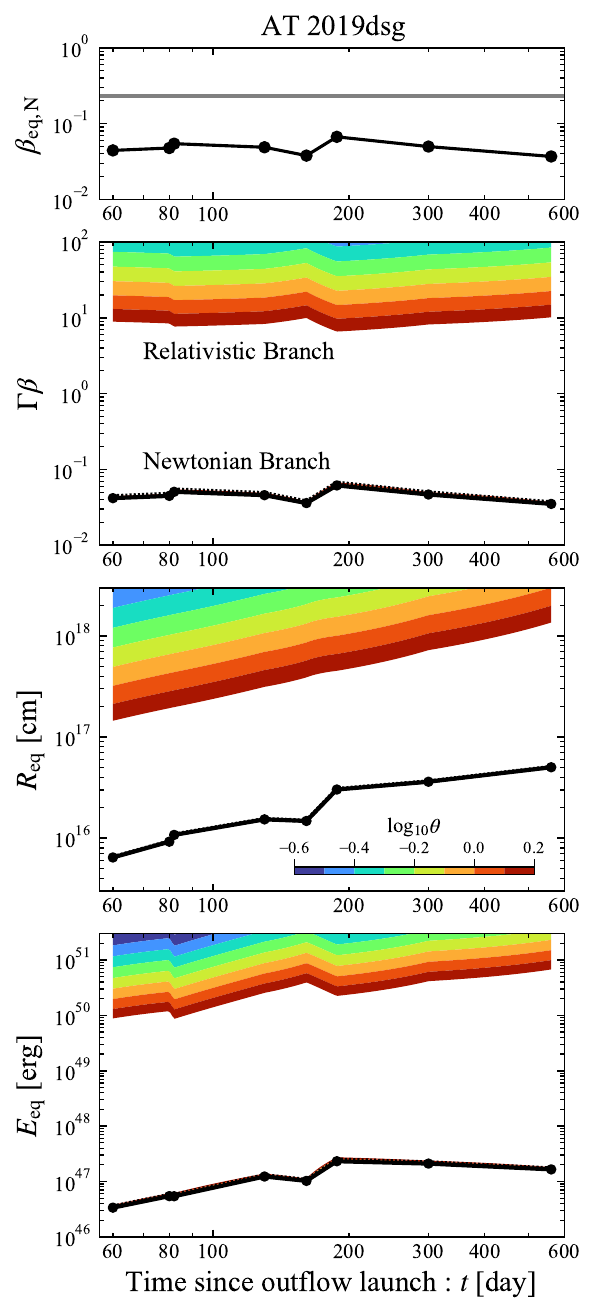}
\caption{The time evolution of the parameter $\beta_{\rm eq,N}$, four-velocity $\Gamma\beta$, and the equipartition radius $R_{\rm eq}$ and energy $E_{\rm eq}$ for AT 2019dsg. The outflow launching time is set 10 days before the discovery. In the top panel, the gray line shows the critical value of $\beta_{\rm eq,N}=0.23$ below which the allowed parameter space splits into Newtonian and relativistic branches. Color contours show the allowed parameter spaces in the second to fourth panels. The black solid and dotted curves show the parameter values for $\theta=0$ and the contour for the boundary between on and off-axis ($\theta=\cos^{-1}\beta$). They almost overlap for small $\beta_{\rm eq,N}$. Observables are linearly interpolated to depict the color map. Note that in the upper regions of all figures, while the angles are very small the solutions are off-axis as $\theta < 1/\Gamma$.}
\label{fig:result_at2019dsg}
\end{center}
\end{figure}

The top panel of Fig.~\ref{fig:result_at2019dsg} depicts the evolution of the parameter $\beta_{\rm eq,N}$ (data is taken from \citealt{Cendes+2021b}). We set the time of the outflow launch as ten days before the optical discovery, as found by previous analyses. Since $\beta_{\rm eq,N}\simeq0.04<0.23$, the possible range of $\Gamma\beta$ has two branches: Newtonian and relativistic velocities. We derive a possible range of four-velocities for each observation epoch, equipartition radii and energies (shown in the bottom panels of Fig.~\ref{fig:result_at2019dsg}). Although the viewing angle has a range of $0\leq\theta\leq\pi$, we show the color contours only up to $\theta\lesssim\pi/2$ (note $\log_{10}(\pi/2)\simeq0.2$) because the outflow is expected to have a bipolar structure such as a jet. We linearly interpolate the observation data between successive observation epochs to depict a continuous color map. As we have discussed in the previous section, $\Gamma\beta$, $R_{\rm eq}$, and $E_{\rm eq}$ attain minimal values for $\theta=0$ in the Newtonian (on-axis) branch. Larger values are needed for different viewing angles. 

The Newtonian branch is essentially the same as the one found in the previous studies \citep{Stein+2021,Cendes+2021b,Matsumoto+2022,Mohan+2022}. For brevity, we did not include in our analysis the energies of the hot proton and total non-thermal electrons (in the case of $\nu_{\rm m}<\nu_{\rm a}$) and a possible deviation from the exact equipartition (see BNP13). However, the radius estimate (hence the Lorentz factor) does not change significantly whether these contributions are included. For example, adding the contribution of hot protons increases the total energy by a factor of $\sim10$ (assuming the non-thermal electron energy is $\sim10\,\%$ of proton energy).

The relativistic branch represents a relativistic jet viewed off-axis disguised as a Newtonian source. With $\beta_{\rm eq,N}\simeq0.04$, the emission is de-boosted by the Doppler factor of $\dop\simeq0.1$ (Eq.~\ref{eq:dop_result}). 
Intriguingly, 
\cite{Mohan+2022}  detected a possible superluminal motion of the radio source at AT 2019dsg: 
$\beta_{\rm VLBI}=3.2\pm2.2$ ($1\sigma$ confidence level). While the result is not statistically significant, we can use these values and estimate (using the intersection of the trajectories of the minimal energy with $\beta_{\rm eq,N}=0.04$ (Eq.~\ref{eq:dop_result}) and the apparent velocity of $\beta_{\rm VLBI}=3.2\pm2.2$  (Eq.~\ref{eq:beta_vlbi})) the corresponding parameters of the system: $\theta\simeq 0.6_{-0.2}^{+0.9}$ and $\Gamma\beta\simeq 50_{-40}^{+100}$ (see Fig.~\ref{fig:veq}). The lowest-energy solution in this branch is realized for $\theta\simeq\pi/2$, $\Gamma\simeq10$, and $E_{\rm eq}$ increasing from $\sim10^{50}\,\rm erg$ to $\sim10^{51}\,\rm erg$. This should be compared with $E_{\rm eq} \simeq 10^{47}\,\rm erg$ for the Newtonian case. While this energy is much larger than the on-axis case, it is reasonably within the total energy budget of a TDE.

However, the expected time evolution of the jet makes this solution unlikely. The jet should eventually decelerate and become Newtonian. Such a transition occurs only if $\beta_{\rm eq,N}>0.23$ for which both branches merge. The current observation finds that the peak flux density already started declining $F_{\rm p}\propto t^{-1.2}$ and the peak frequency decreases $\nu_{\rm p}\propto t^{-1}$, which gives $\beta_{\rm eq,N}\propto t^{-0.6}$ (Eq.~\ref{eq:beta}). Therefore, unless these trends change and either  $F_{\rm p}$ increases or $\nu_{\rm p}$ decreases more rapidly,  $\beta_{\rm eq,N}$ will continue to decrease monotonically, and the transition to the Newtonian branch will never happen.

\subsection{AT 2018hyz}
AT 2018hyz \citep{Gomez+2020,Short+2020,VanVelzen+2021,Hammerstein+2023,Cendes+2022b} shows a relatively bright radio emission $\nu L_\nu \sim10^{39}\,\rm erg\,s^{-1}$ compared to other TDEs, and has a synchrotron self-absorbed spectrum. A peculiar feature is the delayed onset of the radio flare at $\sim 800$ days after the optical discovery \citep[see also][for possible other events]{Horesh+2021,Horesh+2021b,Sfaradi+2022,Perlman+2022}. \cite{Cendes+2022b} carried out an (on-axis) equipartition analysis and rejected the possibility that the outflow is launched at the time of the optical discovery because it requires a non-monotonic velocity evolution. Their preferred solution is that the outflow producing the radio emission was launched 750 days after the discovery with an almost constant velocity $\beta\simeq0.2$ (spherical geometry) or $\beta\simeq0.6$ (jet geometry with $\theta_{\rm j}=10^\circ\simeq0.17$).

\begin{figure}
\begin{center}
\includegraphics[width=85mm, angle=0]{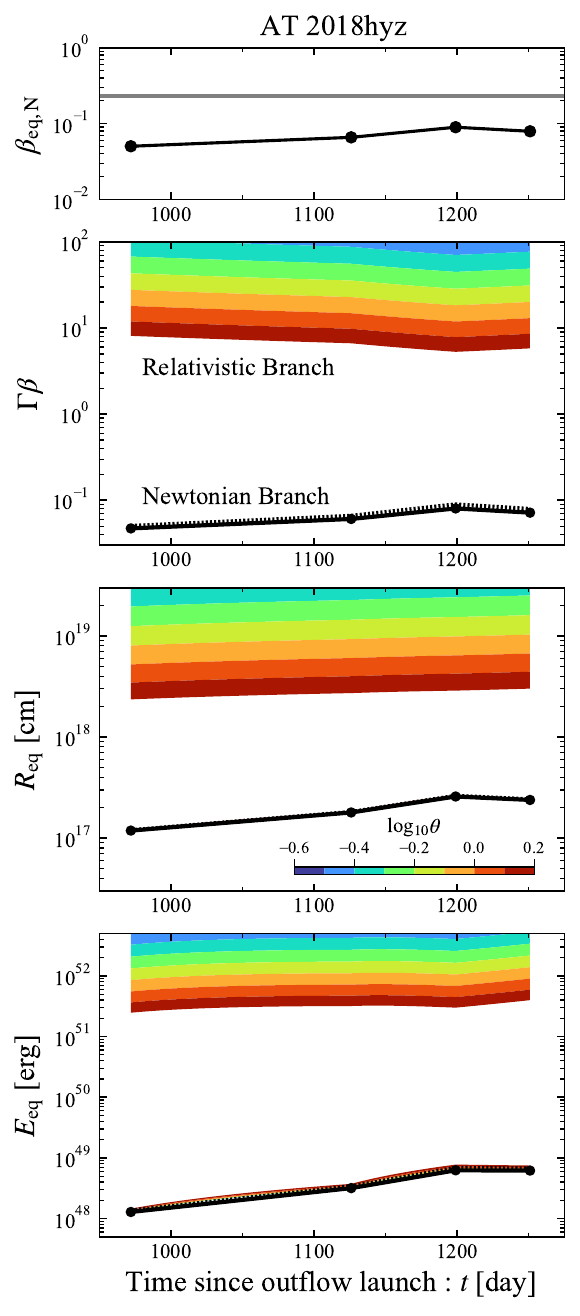}
\caption{The same as Fig.~\ref{fig:result_at2019dsg} but for AT 2018hyz. The outflow launching time is set as the time of discovery.}
\label{fig:result_at2018hyz_t0}
\end{center}
\end{figure}

We examine this event\footnote{Following \cite{Cendes+2022b}, we do not include the data point at 1282 days after the discovery.} assuming that the outflow is launched at approximately the same time as the optical discovery, i.e., roughly at the same time as the TDE. Fig.~\ref{fig:result_at2018hyz_t0} depicts our result. Like in AT 2019dsg, $\beta_{\rm eq,N}<0.23$, and the possible range of quantities is split into two branches. Our results for the Newtonian branch are consistent with those of \cite{Cendes+2022b} except for the correction to the energy as discussed above.

We find that the radio emission may be produced by a relativistic jet viewed off-axis as first studied by \cite{Giannios&Metzger2011}.\footnote{\cite{Cendes+2022b} exclude an off-axis jet arguing that such a scenario predicts a slowly rising flux $F_{\nu}\propto t^{3}$. However, this is not necessarily the case as shown in e.g., \cite{vanEerten+2010,Hotokezaka&Piran2015}.} For a viewing angle of $\theta\simeq\pi/2$ (brown colored stripe in Fig.~\ref{fig:result_at2018hyz_t0}), the outflow has a  slowly decreasing Lorentz factor from $\Gamma\simeq8$ to $\simeq5$ with a Doppler factor of $\dop\simeq0.1$ to $0.2$. The equipartition energy is also weakly increasing from $E_{\rm eq}\simeq2\times10^{51}\,\rm erg$ to $\simeq3\times10^{51}\,\rm erg$. In this solution, an on-axis observer would see a bright emission from this jet. Using the quantities of $\dop\simeq0.2$, $\Gamma\simeq5$, and spectrum at $1251$ days, we estimate the on-axis radio luminosity at 5 GHz of $\nu L_{\nu}\simeq10^{42}\,\rm erg\,s^{-1}$, which would be observed at 20 days after the disruption. This is about ten times larger than the jetted TDE Swift J1644+25 \citep{Zauderer+2011,Berger+2012} but comparable to Swift J2058+05 (\citealt{Brown+2017} and see Fig. 1 in \citealt{Alexander+2020}).

\begin{figure}
\begin{center}
\includegraphics[width=85mm, angle=0]{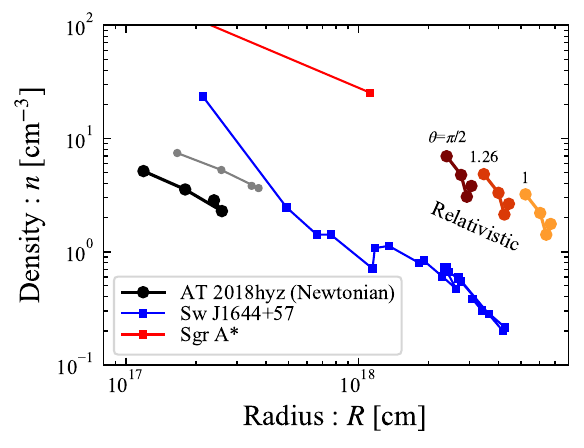}
\caption{CNM density profile reconstructed by our equipartition analysis for AT 2018hyz. The black curve shows the density profile for the on-axis Newtonian branch. The brown to yellow ones show profiles for off-axis relativistic branches with different viewing angles ($\theta=\pi/2$, $1.26(\simeq10^{0.1})$, and $1$ from left to right). The gray and blue curves denote the profiles of AT 2018hyz calculated by \citealt{Cendes+2022b} and of Swift J1644+57 \citealt{Eftekhari+2018}. The red curve shows the profile for our Galactic center \citealt{Baganoff+2003,Gillessen+2019}.}
\label{fig:profile}
\end{center}
\end{figure}

The off-axis jet's energetics is directly estimated using the CNM density. Fig.~\ref{fig:profile} depicts the CNM density profiles reconstructed for both on-axis (Newtonian, black curve) and off-axis (relativistic, brown to orange colored curves) solutions by using Eq.~\eqref{eq:density_on/off}. The density for the on-axis case is consistent with that of \cite{Cendes+2022b} within a factor of 2. The off-axis solution has a density profile about ten times larger than that of the jetted TDE Swift J1644+57 \citep{Eftekhari+2018},\footnote{We multiply a numerical factor of 4 by the densities obtained by \cite{Cendes+2022b,Eftekhari+2018}. They introduced this factor to correct the shock compression, which is, however, not needed to estimate the CNM profile \citep{Matsumoto+2022}.} but it is similar to that of our Galactic center (Sgr A*) if it extends with the same slope. Since the off-axis viewed jet is decelerating, the jet energy is estimated by the total energy of the swept-up CNM. The swept-up mass is obtained by 
\begin{align}
M_{\rm swept}&\sim \MP n\frac{\pi}{\Gamma^2} \frac{R^3}{3}\\
&\simeq3\times10^{-3}\,\Msun\,\left(\frac{n}{3\,\rm cm^{-3}}\right)\left(\frac{R}{3\times10^{18}\,\rm cm}\right)^3\left(\frac{\Gamma}{5}\right)^{-2}\,.
    \nonumber
\end{align}
Therefore, the total kinetic energy of the jet (including the proton's energy) is $E_{\rm kin}\simeq \Gamma^2 M_{\rm swept}c^2\simeq10^{53}\,\rm erg$. The energy of the emitting electron ($E_{\rm eq}\simeq3\times10^{51}\rm erg$) is $\simeq3\,\%$ of the total jet energy. The total jet energy, $10^{53}\,\rm erg$, is about ten times larger than that of jetted TDEs implied by X-ray observations \citep[$\sim10^{52}\,\rm erg$,][]{Bloom+2011,Burrows+2011,Levan+2011,Cenko+2012,Pasham+2015,Brown+2015}, but within the range of the energies required by several radio modelings for Swift 1644+57 \citep[$\sim10^{53}\,\rm erg$,][]{BarniolDuran&Piran2013,Mimica+2015,Generozov+2017}, and recently reported jetted TDE candidate AT 2022cmc \citep{Matsumoto&Metzger2023}.

This off-axis scenario can be tested as $\beta_{\rm eq,N}$ has to increase to allow a late-time transition to the on-axis Newtonian branch within this solution. This requires that either the peak flux increases or the peak frequency decreases. Unlike AT 2019dsg, at the latest observation, the peak flux density of AT 2018hyz is increasing as $F_{\rm p}\propto t^{5}$ (the peak frequency is relatively stable) and hence $\beta_{\rm eq,N}$ increases $\propto t^{1.4}$. Within this model, the peak flux will continue to increase (or peak frequency will start decreasing). Using the evolution of $\Gamma\propto t^{-1.9}$ implied in Fig.~\ref{fig:result_at2018hyz_t0},\footnote{This evolution may be consistent with the evolution of $\Gamma$ after a jet break for an off-axis observer \citep[e.g.,][]{DeColle+2012c}.} we estimate that the jet becomes Newtonian at $\simeq3000$ days after the disruption, and the light curve will peak around that time.

VLBI observations will be able to test the off-axis scenario directly. In the off-axis scenario, the relativistic jet travels roughly perpendicularly to our line of sight at the speed of light. For the distance of $\simeq200\,\rm Mpc$ to AT 2018hyz, such a  source moves at $\simeq 0.3\,\rm mas\,yr^{-1}$ on the sky plane. Given that the radio flux of $\simeq3-10\,\rm mJy$ at the last epoch, which is brighter than the radio afterglow of GRB 170817A, $\sim0.1\,\rm mJy$, for which a superluminal motion was detected \citep{Mooley+2018b,Ghirlanda+2019}, given that the radio signal is rising we expect that  VLBI observation will give an interesting constraint on the off-axis scenario. By the time of the expected peak at $\simeq 3000\,\rm days$, the distance from the image has traversed should be $\simeq 2.4\,\rm mas$. Detection of such a motion will be a ``smoking gun" for the off-axis scenario. Alternatively, a null result will rule it out. If this scenario is confirmed, this jet's implied very large energy will have interesting implications for the central engines in TDEs.

\section{Summary}\label{sec:summary}
We generalized the equipartition method of BNP13 for an arbitrary viewing angle $\theta$. As found by BNP13, the introduction of relativistic motion implies that the total energy inferred from the equipartition analysis can be arbitrarily small (for $\theta=0$). This is because with a higher Lorentz factor, the observed region becomes smaller and smaller, and hence it contains less and less energy. In this sense, the estimated equipartition energy does not reflect the system's actual total energy (presumably broader than $\Gamma^{-1}$). Thus, simply minimizing the energy does not give a solution. 

Following BNP13, we impose another condition relating the radius and the observation time (Eq.~\ref{eq:t} or \ref{eq:r}) to obtain the equipartition radius and energy. Unlike the on-axis scenario, the Lorentz factor cannot be determined uniquely in this case. It becomes degenerate with the viewing angle along the minimal energy trajectory (Eq.~\ref{eq:dop_result}, see also Fig.~\ref{fig:veq}) that is characterized by a velocity parameter $\beta_{\rm eq,N}$. The possible solutions are divided into on-axis and off-axis branches. The latter branch always has larger Lorentz factor and the equipartition radius and energy than those in the on-axis branch. 
For a velocity parameter smaller than a critical value $\beta_{\rm eq,N}\lesssim0.23$, the trajectory is split into these two  (on-axis Newtonian and off-axis relativistic) branches. The off-axis relativistic branch implies that an apparent Newtonian radio source can be a relativistic source viewed off-axis. 

The degeneracy between the Lorentz factor and viewing angle implies that an observed radio signal could have very different interpretations. To demonstrate this, we analyzed two radio TDEs, AT 2019dsg and AT 2018hyz. 
Both are ``apparently'' Newtonian radio TDEs with $\beta_{\rm eq,N}\lesssim0.23$. However, they could be indeed produced by a relativistic jet viewed off-axis. For these objects, long-term monitoring or, equivalently, the evolution of $\beta_{\rm eq,N}$ is critical to distinguish whether they are produced by a Newtonian outflow or relativistic jet. If the radio emitter is a jet and it evolves into the Newtonian branch, $\beta_{\rm eq,N}$ should increase to above the critical value $\beta_{\rm eq,N}\gtrsim0.23$. Current observations of AT 2019dsg find a decreasing velocity parameter, making the possibility of a jet unlikely. 

The situation is different for AT 2018hyz, which involves a delayed radio flare occurring more than two years after the optical discovery. For this event, $\beta_{\rm eq,N}$ is increasing, which suggests that an off-axis relativistic jet is an intriguing alternative explanation for the origin of delayed radio flares. Importantly this scenario can be tested as it predicts an increasing radio flux over the next few years as well as a motion of the radio source that can be measured using VLBI. Further observations of this event, such as exploration of flux and peak luminosity that determines $\beta_{\rm eq,N}$ and VLBI observations of the centroid motion, could distinguish between this solution and the alternative late launch of an outflow. Furthermore, the solution requires a significant jet energy $\sim 10^{53}\,\rm erg$ which is comparable to estimates, based on the radio emission, of the total jet energy of Swift J1644+57 \citep{BarniolDuran&Piran2013,Mimica+2015,Generozov+2017}. This will have remarkable implications for our understanding of TDE's central engines if verified.  

The estimated event rate of jetted TDEs is highly uncertain but consistent with our off-axis scenario for AT
2018hyz. While the rate of on-axis jetted events is poorly constrained, $\sim10^{-2}\,\rm Gpc^{-3}\,yr^{-1}$ \citep{Alexander+2020,DeColle&Lu2020}, taking a  beaming fraction of $f_{\rm beam}\sim10^{-2}$ and total TDE rate of $\sim10^{2}-10^{3}\,\rm Gpc^{-3}\,yr^{-1}$ \citep{VanVelzen2018,Lin+2022}, the fraction of jetted TDE is estimated to be $\sim10^{-3}-10^{-2}$. Since the total number of detected TDE candidates is $\sim 100$ \citep{Gezari2021,Hammerstein+2023,Sazonov+2021}, a few optically detected events can harbor off-axis jets, which are potentially detected as a delayed radio emission. Note that while a jet that points toward us is much easier to detect, it is much more likely that a jet will point elsewhere than toward us.

Our analysis stresses the importance of VLBI observation of late-time radio images of TDEs. Such observations are possibly the best way to reveal the existence of relativistic jets in these systems. The optical signals on which most events are triggered are roughly isotropic, having no preference for on-axis configurations. As such, we will most likely capture TDE jets pointing off-axis away from us. Superluminal motion is a natural consequence if these jets are relativistic.

\section*{acknowledgments}
We thank Kunihito Ioka and an anonymous referee for useful comments. This work is supported in part by JSPS Overseas Research Fellowships (T.M.) and by ERC advanced grants ``TReX'' and ``MultiJets" (T.P.).

\section*{data availability}
The data underlying this article will be shared on reasonable request to the corresponding author.

\bibliographystyle{mnras}
\bibliography{reference_matsumoto}

\appendix
\section{Derivation of peak flux $F_{\rm p}$}\label{sec:append}
We derive Eqs.~\eqref{eq:Fp thin} and \eqref{eq:Fp thick}.
A photon with a frequency $\nu^\prime$ at the rest frame of the emitting region is observed as a photon with a frequency of 
\begin{align}
\nu_{\rm obs}=\frac{\dop\nu^\prime}{1+z}\ ,
	\label{eq:nuobs}
\end{align}
by the Doppler boost and redshift effect.
Let us consider an emitting object whose luminosity distributes over the solid angle at the lab frame by $dL_\nu/d\Omega$.
The observed flux is given by 
\begin{align}
F_{\nu_{\rm obs}}=\frac{(1+z)}{d_{\rm L}^2}\frac{dL_\nu}{d\Omega}\ .
	\label{eq:Fobs}
\end{align}
When the emitting region is optically thin, the luminosity distribution is given by the volume and emissivity:
\begin{align}
\frac{dL_{\nu}}{d\Omega}=j_{\nu}V=\dop^3j_{\nu^\prime}^\prime V^\prime\ ,
	\label{eq:dLdOmega}
\end{align}
where we have transformed the quantities in the lab frame and rest frame, $j_\nu=(\nu/\nu^\prime)^2j_{\nu^\prime}^\prime$, $V=\dop V^\prime$, and $\nu=\dop\nu^\prime$.
Combining Eqs.~\eqref{eq:Fobs} and \eqref{eq:dLdOmega}, the observed flux is given by 
\begin{align}
F_{\nu_{\rm obs}}=\frac{(1+z)\dop^3}{d_{\rm L}^2}j_{\nu^\prime}^\prime V^\prime\ .
	\label{eq:Fnuobsãthin}
\end{align}
This corresponds to Eq. (5.42) of \cite{Dermer&Menon2009}.
By assuming the emission is isotropic at the rest frame and using $j_{\nu^\prime}^\prime V^\prime=P_{\nu^\prime}^\prime N_{\rm e}/4\pi$ where $P_{\nu^\prime}^\prime=\sqrt{3}q_{\rm e}^3B/\ME c^2$ is the synchrotron emissivity, we have Eq.~\eqref{eq:Fp thin}.
When the source is optically thick, the luminosity distribution is given by the intensity and surface area:
\begin{align}
\frac{dL_{\nu}}{d\Omega}=I_\nu A=\dop^3I_{\nu^\prime}^\prime A\ ,
\end{align}
where we used the transformation of $I_\nu=(\nu/\nu^\prime)^3I_{\nu^\prime}^\prime$.
Finally, we have the observed flux of 
\begin{align}
F_{\nu_{\rm obs}}=\frac{(1+z)\dop^3}{d_{\rm L}^2}I_{\nu^\prime}^\prime A\ .
	\label{eq:Fnuobs thick}
\end{align}
When the system is optically thick, the intensity is given by the Planck function. In particular, if the observed frequency is smaller than the peak frequency, the Rayleigh-Jeans formula gives Eq.~\eqref{eq:Fp thick}.

\section{Narrow Jet geometry}\label{sec:append2}
\begin{figure}
\begin{center}
\includegraphics[width=85mm, angle=0]{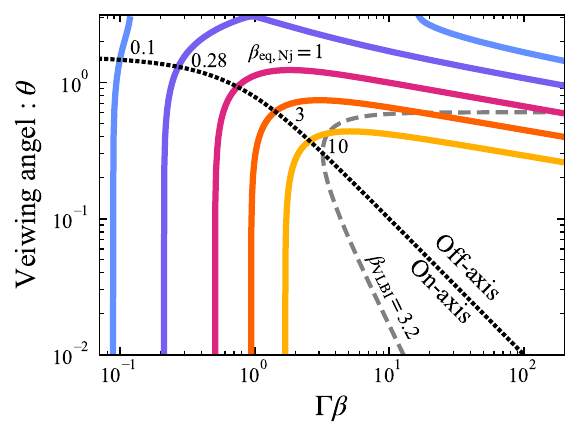}
\caption{The same as Fig.~\ref{fig:veq} but for a narrow jet geometry.}
\label{fig:veq_j}
\end{center}
\end{figure}

\begin{figure}
\begin{center}
\includegraphics[width=85mm, angle=0]{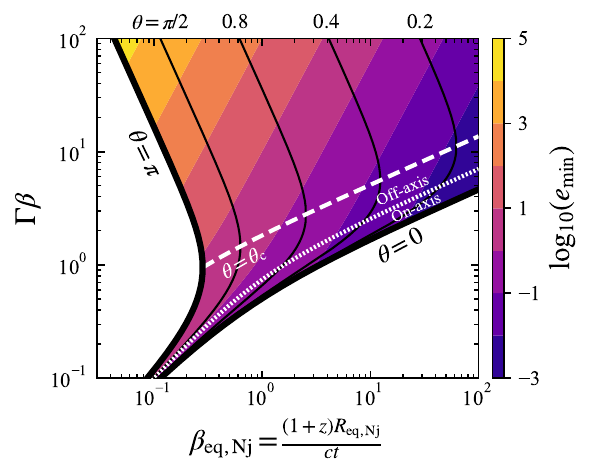}
\caption{The same as Fig.~\ref{fig:veq2} but for a narrow jet geometry.}
\label{fig:veq2_j}
\end{center}
\end{figure}

If a jet can keep its half-opening angle $\theta_{\rm j}$ even after a significant deceleration $\Gamma\ll1/\theta_{\rm j}$, we have to take into account the geometrical effect by modifying the factors of 
\begin{align}
f_{\rm A}=f_{\rm V}=(\Gamma\theta_{\rm j})^2\ .
\end{align}
The procedure to obtain the equipartition quantities is the same as that in the case of $f_{\rm A}=f_{\rm V}=1$, and we have
\begin{align}
R_{\rm eq}&= R_{\rm eq,Nj}\Gamma^{1/17}\dop^{-7/17}\ ,\\
R_{\rm eq,Nj}&\simeq1.6\times10^{18}{\,\rm cm\,}\left[\frac{F_{\rm p,mJy}^{\frac{8}{17}}d_{\rm L,28}^{\frac{16}{17}}\eta^{\frac{35}{51}}}{\nu_{\rm p,10}(1+z)^{\frac{25}{17}}}\right]\left(\frac{\theta_{\rm j}}{0.1}\right)^{-16/17}\ ,
	\nonumber\\
E_{\rm eq}&= E_{\rm eq,Nj}\Gamma^{11/17}\dop^{-43/17}\ ,\\
E_{\rm eq,Nj}&\simeq1.4\times10^{50}{\,\rm erg\,}\left[\frac{F_{\rm p,mJy}^{\frac{20}{17}}d_{\rm L,28}^{\frac{40}{17}}\eta^{\frac{15}{17}}}{\nu_{\rm p,10}(1+z)^{\frac{37}{17}}}\right]\left(\frac{\theta_{\rm j}}{0.1}\right)^{-6/17}\ .
	\nonumber
\end{align}
Here the Newtonian radius $R_{\rm eq,Nj}$ and energy $E_{\rm eq,Nj}$ are given by replacing $f_{\rm A}$ and $f_{\rm V}$ with $\theta_{\rm j}^2$ in the original equations (Eqs.~\ref{eq:Req} and \ref{eq:Eeq}). Now the energy (Eq.~\ref{eq:E2}) and minimal energy trajectory (Eq.~\ref{eq:dop_result}) are modified by
\begin{align}
e(r,\Gamma,\theta)&=\Gamma^{\frac{11}{17}}\dop^{-\frac{43}{17}}\left[\frac{11}{17}\left(\frac{r}{\Gamma^{\frac{1}{17}}\dop^{-\frac{7}{17}}}\right)^{-6}+\frac{6}{17}\left(\frac{r}{\Gamma^{\frac{1}{17}}\dop^{-\frac{7}{17}}}\right)^{11}\right]\ ,\\	\dop&=\left(\frac{\beta_{\rm eq,Nj}}{\beta}\right)^{17/24}\Gamma^{-2/3}\ ,
\end{align}
where
\begin{align}
\beta_{\rm eq,Nj}&\equiv\frac{(1+z)R_{\rm eq,Nj}}{ct}\\
&\simeq6.3\,\left[\frac{F_{\rm p,mJy}^{\frac{8}{17}}d_{\rm L,28}^{\frac{16}{17}}\eta^{\frac{35}{51}}}{\nu_{\rm p,10}(1+z)^{\frac{8}{17}}}\left(\frac{t}{100{\,\rm day}}\right)^{-1}\right]\left(\frac{\theta_{\rm j}}{0.1}\right)^{-16/17}\ .
    \nonumber
\end{align}
Figs.~\ref{fig:veq_j} and \ref{fig:veq2_j} show the same figures as Figs.~\ref{fig:veq} and \ref{fig:veq2} but for a narrow jet geometry. The qualitative feature does not change.

\label{lastpage}
\end{document}